\documentstyle[preprint,aps,epsfig]{revtex}

\begin{document}


\draft
\preprint{
\vbox{
\hbox{ADP-98-11/T289}
\hbox{IU/NTC 98-03}
}}

\title{Shadowing in neutrino deep inelastic scattering
and the determination of the strange quark distribution}
\normalsize 
\author{C.Boros$^1$, J.T.Londergan$^2$ and A.W.Thomas$^1$}
\address {$^1$ Department of Physics and Mathematical Physics,
                and Special Research Center for the
                Subatomic Structure of Matter,
                University of Adelaide,
                Adelaide 5005, Australia}
\address{$^2$ Department of Physics and Nuclear
            Theory Center, Indiana University,
            Bloomington, IN 47404, USA}

\date{\today} 
\maketitle

\begin{abstract}  
We discuss  shadowing corrections to the structure function $F_2$ in 
neutrino deep-inelastic scattering on heavy nuclear targets. 
In particular, we  examine  
the role played by shadowing in the comparison  of the  
structure functions $F_2$ measured in neutrino and muon deep inelastic 
scattering. The importance of  shadowing corrections 
in the determination of the strange quark 
distributions is explained.   
\end{abstract}

\vspace*{3cm}

\vfill\eject

\tighten

\section{Introduction}

Comparisons of  structure functions measured in different 
reactions have always been very useful in investigating    
the structure of hadrons and extracting 
the parton distribution functions.  Recently  
there has been much interest in the measurement of the structure 
function  $F_2^{\nu}(x,Q^2)$  
 in neutrino deep inelastic scattering by the CCFR-Collaboration 
\cite{CCFR}.  
This measurement makes it possible to compare structure functions 
measured in neutrino-induced reactions  
with those measured in charged lepton-induced ones and hence   
to test the universality of parton distribution 
functions and to extract the strange quark density of the nucleon.     

The CCFR-Collaboration compared the neutrino structure function 
$F_2^\nu (x,Q^2)$ extracted from their data on an iron target \cite{CCFR}   
with $F_2^\mu (x,Q^2)$ measured for the deuteron by the NMC 
Collaboration \cite{NMC}.  In the region of intermediate values of
Bjorken $x$ ($0.1 \le x \le 0.4$), they found very good agreement between 
the two structure functions.  In the small 
$x$-region however ($x < 0.1$), the CCFR group found that the two 
structure functions differ by as much as 10-15$\%$. Since several 
corrections have to be taken into account in order to compare the
structure functions  $F_2^\nu (x,Q^2)$ and  $F_2^\mu (x,Q^2)$, 
the apparent discrepancy between the structure functions at small $x$ 
depends on the validity of the assumptions made in correcting 
the data. One of the crucial points is that the neutrino structure 
function is measured on an iron target while the muon data is taken on 
the deuteron. Thus, one has to account for heavy target effects in the
neutrino reactions. In applying these corrections to the data it has 
been assumed that heavy target effects are the same in neutrino and muon 
deep inelastic scattering, and a parametrization obtained from 
muon data has been used. 

{\it A priori} there is no reason {\it why} heavy target corrections in 
neutrino deep inelastic scattering should be the same as those in 
charged lepton deep inelastic scattering.  Therefore, we feel that it is 
important to investigate the role played by shadowing in neutrino
reactions {\it before} concluding that the two structure functions  
are {\it really} different in the small $x$-region. Furthermore, there 
are additional uncertainties arising because the heavy target    
corrections are applied by parametrizing {\it only} the $x$-dependence 
of the available data on the ratio 
$R\equiv F^{\ell A}_{2}(x,Q^2)/F^{\ell D}_{2}(x,Q^2)$ between the 
neutrino structure function measured on heavy targets 
and that of the deuteron from charged lepton deep inelastic scattering.  
However, it is well known that shadowing corrections are very much 
$Q^2$ dependent for smaller $Q^2$ values (where a considerable part 
of the available data was taken), and the $Q^2$ and $x$-dependence of 
the data are strongly correlated because of the 
fixed target nature of these experiments.   

In view of these uncertainties, the main objective of this paper is  
a careful re-analysis of the shadowing corrections which must be 
understood before one can attribute the discrepancy between 
$F_2^{\nu}(x,Q^2)$ and $F_2^{\ell}(x,Q^2)$ to other possibilities, 
such as to different strange quark and anti-strange quark distributions 
\cite{ST,Brodsky,sdiff,JT,HSS}, to higher order QCD-corrections 
\cite{Tung,Barone,Reya} or to the violation of charge symmetry 
in parton distribution functions \cite{Lond,Sather,Lon98,Ben97,Ben98}.

\section{Comparison of neutrino and muon structure functions}

Comparisons of structure functions measured 
in neutrino deep-inelastic scattering with those 
measured in charged lepton deep-inelastic scattering are based on 
the interpretation of these structure functions in terms of 
parton distribution functions in the quark parton model. 
Assuming the validity of charge symmetry and neglecting  
the contributions from charm quarks,  
the structure functions $F_2^{\nu N_0}(x,Q^2)$ and 
$F_2^{\ell N_0}(x,Q^2)$ on iso-scalar targets ($N_0$) 
are given by the following expressions: 
\begin{eqnarray}
  F_2^{\ell N_0}(x) & =& \frac{5}{18} x[ u(x) + \bar u(x)
 +d(x) +\bar d(x) + \frac{2}{5} (s(x) + \bar s(x))] \\
  F_2^{\nu N_0} (x) &=& x[u(x)+ \bar u(x) +d(x) +\bar d(x)
     + 2 s(x)].  
\end{eqnarray} 
Thus, they can be  related to each other  by 
\begin{equation} 
  F_2^{\ell N_0}(x,Q^2)= \frac{5}{18} 
F_2^{\nu N_0 }(x,Q^2) - \frac{3x[s(x)+\bar s(x)]+5x[s(x)-\bar s(x)]} 
{18}. 
\label{eq:1}
\end{equation}
This means that, once the charged lepton and neutrino structure functions  
and the strange quark distributions are known, one can test the validity 
of this relation, or one can use the above relation to extract the 
strange quark distribution from the measured structure functions.

The recent measurement of the structure function  
$F_2^\nu$ by the CCFR-Collaboration \cite{CCFR} makes it possible 
to carry out such an analysis for the first time with reasonable 
precision.  However, the actual comparison between neutrino and charged  
lepton structure functions is not straightforward because 
several corrections have to be applied to the data.   
Since the above relations are only valid 
for $Q^2$ values well above charm  production threshold,  
charm threshold effects have to be removed. Furthermore, 
the neutrino structure function has been extracted from measurements 
using an iron target. Therefore one has to account for 
the excess of neutrons in iron (iso-scalar corrections) and 
for heavy target effects. 

In applying the heavy target corrections one could assume that heavy 
target effects are the same in {\it neutrino} deep inelastic scattering
as in {\it muon} deep inelastic scattering and use a  
parametrization of the heavy target corrections obtained from 
muon-induced reactions. This is the assumption which has been 
made by the CCFR-Collaboration in its analysis \cite{CCFR}.      
Using such a parametrization for the heavy target correction and 
a parametrization of the strange quark distribution \cite{Lai}  
extracted from other experiments, we calculated the ``charge 
ratio'': 
\begin{equation}
 R_c(x)\equiv \frac{F_2^{\mu N_0}(x)}{\frac{5}{18}
 F_2^{\nu N_0}(x) -\frac{x(s(x)+\bar s(x))}{6} } 
\approx 1 - \frac{s(x)-\bar s(x)}{Q_s(x)}.  
\label{Rc}  
\end{equation} 
Here, we defined $Q_s(x) \equiv \sum_{q=u,d,s} [q(x)+\bar q(x)] - 
3(s(x)+\bar{s}(x))/5$. For the charged lepton structure function we used 
 $F_2^{\mu N_0}$ measured in muon deep inelastic scattering 
by the NMC-Collaboration on a deuteron target \cite{NMC}.    
For fixed $x$-values we averaged the structure function over the 
overlapping $Q^2$-regions of the two experiments, in order to obtain 
better statistics. We also applied 
a cut for $Q^2$ less than $3.2$ GeV$^2$ and 
$2.5$ GeV$^2$ for the CCFR data and the NMC data repectively,   
in order to insure the validity of quark-parton model relations.  
For the strange quark distributions we used the CTEQ (CTEQ4L) distributions
of Lai {\it et al.} \cite{Lai}.  

\begin{figure}
\epsfig{figure=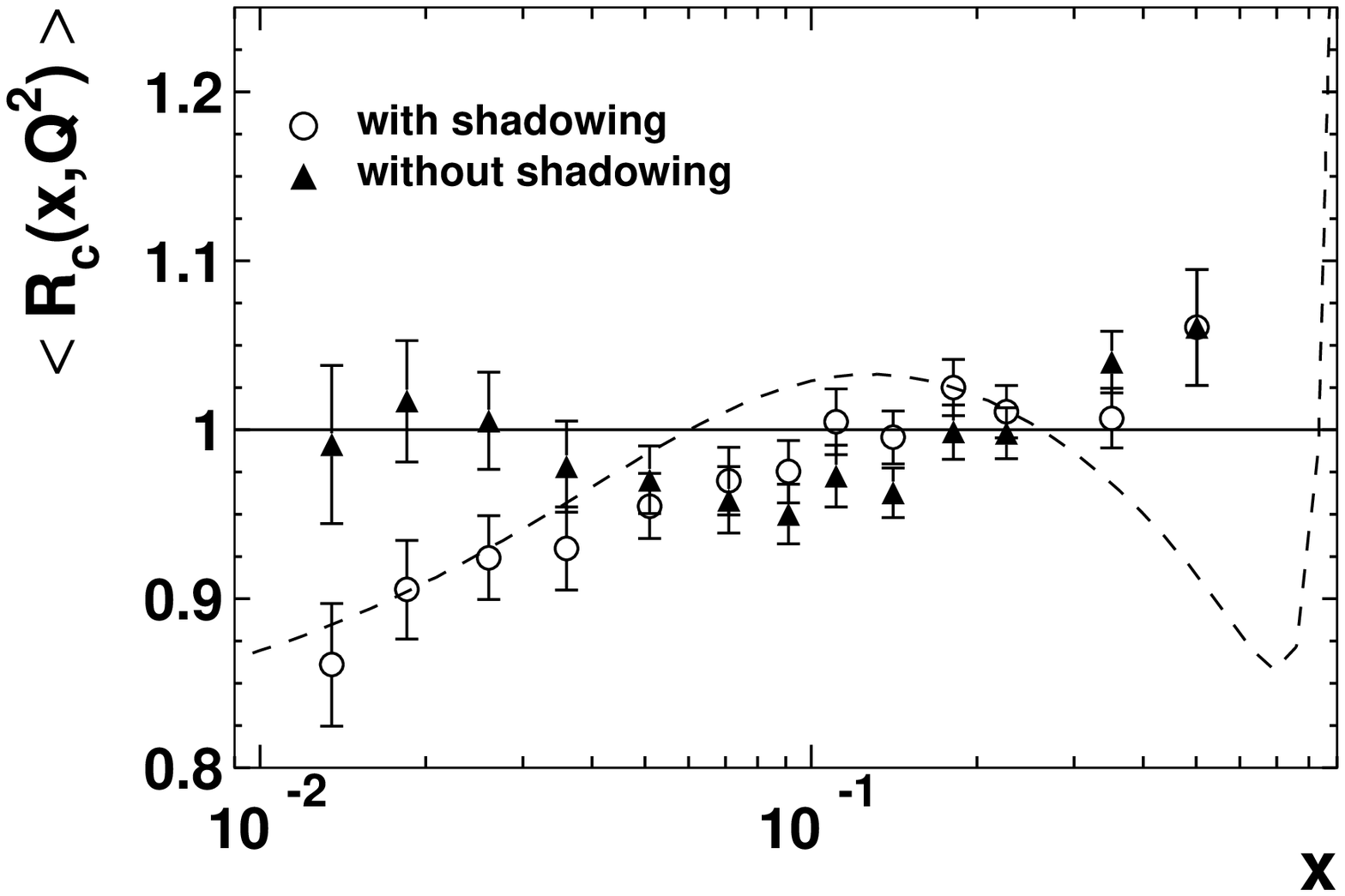,height=12.cm} 
\caption{The ``charge ratio'' $R_c$ of Eq.\ \protect\ref{Rc} 
	as a function of $x$ calculated using 
         the CCFR \protect\cite{CCFR} data for neutrino and 
         NMC \protect\cite{NMC} data for muon 
         structure functions. The data have been integrated 
         over the overlapping kinematical regions 
         and  have been corrected for heavy target effects 
         using a parametrization (dashed line) for heavy target corrections 
         extracted from charged lepton scattering. The result is 
         shown as open circles.  
         The ratio  obtained without heavy target corrections 
         is shown as solid triangles.  
         Statistical and systematic errors are added in 
         quadrature. }
\label{fig1} 
\end{figure}

The result is shown in Fig.\ref{fig1}. 
We note that, under the assumptions that $s(x)=\bar s(x)$ and that 
charge symmetry is valid for parton distributions, the  ``charge ratio'' 
$R_c$ of Eq.\ \ref{Rc} should be equal to one at all $x$.  
For intermediate values of Bjorken $x$ ($0.1 \le x \le 0.4$), the 
charge ratio $R_c$ is equal to one to within errors of a few percent.  
The agreement between the two structure functions in this $x$ region 
allows us to place rather strong upper limits on contributions from charge 
symmetry violation in parton distributions \cite{Lon98}.  However,  
$R_c$ appears to be substantially below unity in the small-$x$ region, 
for $x < 0.1$.  

In Fig.\ref{fig1} we also show the effects of the heavy target 
corrections which were applied to the neutrino structure functions.  
The solid triangles show the result we would obtain for the ``charge 
ratio'' if we did not apply any heavy target corrections to the 
neutrino structure functions.  We see that the heavy target corrections 
definitely play a very important role in interpreting the result of 
such an analysis.   Since the heavy target corrections applied
to the neutrino results were obtained from data in 
charged lepton deep-inelastic scattering, differences between 
shadowing for neutrinos and for muons could make a substantial 
difference in the charge ratio $R_c$ in Eq.\ \ref{Rc}.       
Since the heavy target corrections for large $x$-values are expected to 
be independent of the probe used to measure the quark distributions 
in a nucleus (at large $x$ the target corrections should be dominated
by quark Fermi motion), in this paper we discuss only the shadowing 
region, $x\le 0.1$.



\section{Shadowing corrections}

In calculating the shadowing corrections we use a two-phase model 
which has been  successfully applied to the description of  shadowing 
 in charged-lepton deep inelastic scattering \cite{Badelek,Melni}.  
This approach uses vector meson dominance (VMD)  to describe 
the low-$Q^2$, virtual photon interactions, and Pomeron exchange 
for the approximate scaling region. It is ideally suited    
to describe the transition region between large-$Q^2$ and   
small-$Q^2$.  This is the kinematic region where the largest 
differences occur between the NMC and CCFR data sets.

First, we discuss hadron dominance  for neutrino deep 
inelastic scattering \cite{Stodolsky,VMD}.  
The basic physical picture   
is that the photon or vector boson fluctuates into
a quark-antiquark pair before interacting with the nucleus.
If the life-time of such a fluctuation is long enough a coherent
hadronic state can build up before interacting  with the target, 
leading to shadowing characteristic of hadrons \cite{VMD,Bell}.    
To generalize VMD to neutrino scattering we have to include 
both pseudo-scalar mesons (pions) and  
axial vector mesons ($A_1$..),  because of the 
(V-A) nature of the weak currents.  

In order to identify the contributions of the different, virtual 
hadronic states to the nucleon structure functions,  
we note that the hadronic tensor for deep inelastic 
neutrino scattering is defined by:
\begin{equation}
 W_{\mu\nu}(\nu,Q^2)= \frac{1}{2} \sum_{S}\sum_{X} 
 \langle PS | J_\mu | X\rangle \langle X | J_\nu| PS \rangle 
(2\pi)^3 \delta^4(P+q-p_X)
\label{Wuv} 
\end{equation} 
In Eq.\ \ref{Wuv}, $q_\nu$ and $\nu$ are the momentum and energy transfer 
from the neutrino to the nucleon; $Q^2=-q^2$ is the invariant mass 
of the $W$-boson; $M$, $P$ and $S$ are the mass, four-momentum 
and spin of the target nucleon; $p_X$ is the four-momentum of the final 
state $X$.   
$J_\mu=V_\mu - A_\mu$ is the weak current with vector ($V_\mu$) 
and axial vector ($A_\mu$) components respectively. 
$W_{\mu\nu}$ can be parametrized in terms of six invariant structure 
functions $W_i(\nu,Q^2)$ in the following form: 
\begin{eqnarray} 
\frac{1}{2M} W_{\mu\nu }(\nu,q^2)
&= &  -g_{\mu\nu} W_1(\nu,q^2)
+\frac{P_\mu P_\nu}{M^2} W_2(\nu,q^2) - 
\frac{i\epsilon_{\mu\nu\alpha\beta} 
P^\alpha q^\beta}{2M^2} W_3(\nu,q^2)+ \nonumber\\  
&+& \frac{q_\mu q_\nu}{M^2} W_4(\nu,q^2) 
+ \frac{P_\mu q_\nu + P_\nu q_\mu}{2M^2} W_5(\nu,q^2) 
+ i\frac{P_\mu q_\nu - P_\nu q_\mu}{2M^2} W_6(\nu,q^2) 
\label{eq:tensor} 
\end{eqnarray}
In contrast to the vector current, the axial current is not conserved. 
Thus, we cannot impose current conservation on 
$W_{\mu\nu}$. Therefore, in addition to the three structure functions
which arise for vector currents, another three structure functions 
appear in  Eq.\ (\ref{eq:tensor}) for the axial current.  
In the following we are interested only in the 
symmetric, parity conserving piece of the hadronic tensor 
and want to discuss the major differences 
between axial and vector currents which are relevant to this work. 
(More detailed discussions can be found in Refs.
\cite{Stodolsky,VMD,Boris1}).

Hadronic dominance assumes that the weak current is dominated 
by intermediate hadronic states coupled to the weak current.  
The generalization of  vector meson dominance to axial 
vector mesons is straightforward. Here, we quote only the result 
\cite{Stodolsky}.  
The contribution of the  vector mesons and axial vector mesons 
 to the structure function 
$F_2(x,Q^2)= \nu W_2(\nu,Q^2)$ can be written in the familiar 
form: 
\begin{equation}
 F^{VMD}_{2}(x,Q^2)=  
\frac{Q^2}{\pi} \sum_{V=\rho^+,A_1...} (\frac{f_V}{Q^2+m^2_V})^2
 \sigma_{VN}   
\label{vmdeq} 
\end{equation}
Here,  $f_V$  are the vector meson  
coupling constants; $m_V$ 
are the masses of the vector mesons, $\sigma_{VN}$  
is the vector meson target total cross section. 
Since the vector mesons couple differently to the electromagnetic 
and to the weak current, the coupling constants 
are different in neutrino and charged lepton scattering. 
Their relative strength can be determined according to the 
quark counting rules \cite{Stodolsky}. It turns out that, once the overall 
weak and electromagnetic coupling constant is removed,  
the relative coupling of $\rho^+$ and $A_1$ to the $W$-boson 
($f^2_{\rho^+}=f^2_{A_1}$)  
is twice as large as the coupling of the $\rho^0$ to the photon,  
$f^2_{\rho^0}$.

The main difference between the axial and vector currents is related to 
the fact that axial currents are only partially conserved (PCAC).  
Adler's theorem \cite{Adler} 
relates  the divergence of the axial current to the pion field $\Phi$, 
for $Q^2=0$ 
\begin{equation} 
    \partial_\mu A^\mu =f_{\pi}\, m_\pi^2 \, \Phi, 
\end{equation} 
where $f_\pi=0.93\, m_\pi$ is the pion decay constant and $m_\pi$ 
the pion mass.   
Imposing this constraint on the hadronic tensor, $W_{\mu\nu}$,  
we see that only the term containing $W_2$ survives the limit 
$Q^2\rightarrow 0$ and we obtain the following 
contribution from PCAC to the structure function $F_2(x,Q^2 )$: 
\begin{equation}
  F_2^\pi(x,Q^2) = \frac{f^2_{\pi}}{\pi} \sigma_{\pi N},   
\end{equation}
where $\sigma_{\pi N}$ is the pion nucleon total 
cross section. 
However, it is important to note that this is not 
a consequence of the pion dominance of axial currents. 
In order to see this we write  the 
(matrix element of the) axial vector 
current in terms of the pion-pole term 
\begin{equation} 
  A_\mu = A^{\prime}_\mu + f_\pi \frac{q_\mu}{Q^2+m_\pi^2} 
T^{\pi N \rightarrow X} 
\label{pionint}
\end{equation} 
Here, the second term stands for the contribution of the pion-pole, 
with $T^{\pi N\rightarrow X}$ being the $\pi N\rightarrow X$ transition 
amplitude, 
and $ A^{\prime}_\mu $ 
the other contributions; for example  
the contribution from axial vector mesons. 
Now comparing this expression with the hadronic tensor, Eq. 
(\ref{eq:tensor}),  we immediately see 
that the pion-pole  and its interference terms with 
$A^{\prime}$ will only contribute 
to the structure functions $W_4$ and $W_5$ 
but not to the structure function $W_2$. Besides, 
the pionic contributions to the cross section will be proportional 
to the mass of the outgoing muon, $m_\mu$, because the leptonic tensor 
is conserved up to terms proportional to $m_\mu$. Thus, the coupling 
of virtual pions to the axial current is strongly suppressed.

This is a remarkable result.  Although the axial current cannot ``emit 
a pion in the vacuum'' \cite{Bell}, the cross section for neutrino 
scattering on a nucleon is proportional to the pion cross section on 
the same target. The observation that 
the PCAC-term is not to be attributed to the pion-pole, but rather 
to the longitudinal component of higher mass terms ($A_1..$) 
\cite{Stodolsky},  helps to resolve the apparent contradiction.  
PCAC thus provides a relation between the higher mass contributions 
to the axial current and the pion cross section.  
If we identify  the PCAC component with the 
longitudinal part of the $A_1$  we have the following 
constraint for the longitudinal cross section: 
\begin{equation}
 \sigma^{A_1 N}_L = \frac{1}{Q^2} \frac{f^2_{\pi}}{f_{A_1}^2}
m_{A_1}^4 \sigma_{\pi N}
\label{longitudinal}
\end{equation}
Inserting Eq.(\ref{longitudinal}) back  in Eq.(\ref{vmdeq}) 
we obtain our final expression 
for the PCAC-term:
\begin{equation}
  F_2^\pi(x,Q^2) = (\frac{m_{A_1}^2}{Q^2+m_{A_1}^2})^2
\frac{f_{\pi}^2}{\pi} \sigma_{\pi N} 
\label{pion}
\end{equation}
Since, experimentally, relation (\ref{longitudinal}) 
does not hold with the $A_1$ alone, one should  
 include other higher mass contributions. In fact, one should 
integrate over the whole diffractively produced spectrum, as was pointed 
out in Ref.\cite{Boris1}. However, if such an integration is
performed, Eq.\ (\ref{pion}) remains a very good approximation for 
small $Q^2$-values with a mass  which is not exactly the 
same but is extremely close to $m_{A_1}$ \cite{Boris1}.  
The presence of the pion-term for small $Q^2$ is experimentally 
well established. Experiments on diffractive meson-production  
\cite{Diffmeson}, on total cross sections \cite{Total} in neutrino- 
and antineutrino interactions and shadowing \cite{Shadow}  in neutrino 
deep inelastic scattering for very small $Q^2$ values,  
 have confirmed the validity  of PCAC \cite{Adler}. 
 
Finally, it should be noted here that the non-vanishing of the 
longitudinal cross section  for $Q^2\rightarrow 0$ 
in  neutrino deep inelastic scattering, as a consequence of PCAC, leads 
to a ratio $R\equiv \sigma_L/\sigma_T$ which is different from      
that in muon deep inelastic scattering where current 
conservation (for vector currents) forces $\sigma_L\rightarrow 0$ 
in the $Q^2\rightarrow 0$ limit. In the extraction 
of the structure function by the CCFR-Collaboration it was assumed 
that $R$ is the same in both processes.  However, this assumption 
has little effect on the charge ratio, since the pionic 
contribution does not play a significant role in the kinematic region of 
the CCFR experiment which is the focus of this paper.    

The vector meson undergoes multiple scattering while    
traveling through nuclear matter.     
The resulting shadowing can be calculated using  
the Glauber multiple scattering expansion \cite{Glauber}.    
In the eikonal approximation this gives the following correction to 
the nuclear structure function \cite{Badelek,Melni}: 
\begin{equation}
A\delta^{(V)} F^{\nu A}_{2}(x,Q^2) =
 \frac{Q^2}{\pi} \sum_{\rho^+,A_1..}
 \frac{f_V^2}
{(Q^2+m_V^2)^2} \delta\sigma_{VA}
\end{equation}
where
\begin{eqnarray}
  \delta\sigma_{VA}& =& -\frac{1}{2} A(A-1)\, \, \sigma^2_{VN}
  \mbox{Re} \int_{z'>z}d^2b\, dz \,dz'\,\, \mbox{exp}[ik^V_L(z'-z)]
\\ \nonumber  & \times & \rho^{(2)}(\vec{b},z,z')\, \mbox{exp}
(-\frac{A}{2}\int_z^{z'}
\frac{d\xi}{L_V})
\end{eqnarray}
is the shadowing correction to the vector meson-nucleus 
cross section with impact parameter $\vec{b}$ and 
longitudinal momentum transfer to the nucleon $k^V_L=Mx(1+m_V^2/Q^2)$. 
$M$, $m_V$ and $f_V$ are the nucleon mass, vector meson masses and 
vector meson coupling constants, respectively. Further,  
$\rho^{(2)}(\vec{b},z,z')=N_c \rho (\vec{r})\rho  (\vec{r'})$   
is the two-body density function, normalized 
according to $\int d^3r d^3r' \rho^{(2)}(\vec{r},\vec{r'}) 
=\int d^3r \rho (\vec{r})=1$. For the single particle density in  
iron we use the Woods-Saxon (or Fermi) density with typical parameters 
given in Ref.\ \cite{Jackson}. The mean 
free path of the vector mesons in the nucleus, $L_V$, is given   
by $L_V=[\sigma_{VN} \rho(\vec{b},\xi )]^{-1}$. 
For the total vector meson cross sections we use the energy 
dependent parametrizations of Donnachie and Landshoff \cite{Lands},   
$\sigma_{\rho N}=\sigma_{A_1 N}=13.63 s^\epsilon+31.79 s^{-\eta}$, 
where $s$ is the 
photon-nucleon total centre of mass (CMS) energy, 
 $s=(P+q)^2$, with $P$ and $q$ the 
four-momenta of the nucleon and photon, respectively. The parameters 
$\epsilon\approx 0.08$ and $\eta\approx 0.45$ are motivated by Regge 
theory. Finally, 
the relative strength of the coupling constants can be determined
according to the quark counting rule:
$f^2_{\rho^+}:f^2_{A_1}:f^2_{\rho^0}= 1:1:\frac{1}{2}$.  
We use the experimental values for the coupling constants in
charged lepton scattering ($f_V\equiv M^2_V/\gamma_V$ and
$\gamma_V^2/4\pi =2.0$, $23.1$, $13.2$ for $V=\rho^0$, $\omega$,
$\phi$ \cite{VMD})
and calculate the coupling of the weak current to the
$\rho^+$ and $A_1^+$ according to the above relation.

The pionic component, arising through  
PCAC, will be shadowed in the same way as 
the vector meson components \cite{Bell}: 
\begin{equation}
A\delta^{(\pi)} F^{\nu A}_{2}(x,Q^2) =
 \frac{f^2_{\pi}}{\pi}
(\frac{m^2_{A_1}}{Q^2+m^2_{A_1}})^2
 \delta\sigma_{\pi A}
\end{equation}
where
\begin{eqnarray}
  \delta\sigma_{\pi A}& =& -\frac{1}{2} A(A-1)\, \, \sigma_{\pi N}^2
  \mbox{Re} \int_{z'>z}d^2b\, dz \,dz'\,\, 
\mbox{exp}[ik^\pi_L(z'-z)]\nonumber \\
 &\times & \rho^{(2)}(\vec{b},z,z')\, \mbox{exp}
(-\frac{A}{2}\int_z^{z'}
\frac{d\xi}{L_\pi}), 
\label{eq:pion}
\end{eqnarray}
is the shadowing correction to the pion-nucleon total cross section. 
For  the pion-nucleon total cross section we use 
$\sigma_{\pi N}=24$ $mbarn$ and  for the pion decay constant 
$f^2_\pi=(0.93 m_\pi)^2$ \cite{Diffmeson}.    
Here we note  the following. The appearance of the pion mass 
in $k^\pi_L$  has the effect that shadowing in neutrino 
scattering (at $Q^2\approx 0$) sets in and saturates 
at much lower energies than in charged lepton 
scattering, where the coherence condition is governed by the 
$\rho$-mass. The early onset of shadowing at low energies 
has been confirmed experimentally by the BEBC-Collaboration \cite{Shadow} 
and suggests that in neutrino scattering at low $Q^2$ it is 
the pion which propagates through the nuclear medium and leads to 
shadowing. This experimental fact is to be compared to the 
observation that axial currents cannot emit pions in the vacuum as mentioned 
above. However, in nuclear medium 
pions can be diffractively produced as pointed out by Kopeliovich 
\cite{Boris2}. According to this interpretation  
one should also take into account contributions from inelastic 
shadowing arising from  diffractive dissociation 
of the pions \cite{Boris2}.   
Since this inelastic shadowing gives only small corrections to the 
elastic pion contribution, Eq.\ (\ref{eq:pion}), and  
the inclusion of the pion component is only important 
for small $Q^2\sim m^2_\pi$ and 
negligible for $Q^2 \ge 1$ GeV$^2$, 
in the following we neglect the tiny contributions from inelastic 
pion shadowing. 

While shadowing due to VMD and PCAC dominates for small $Q^2$-values,
at high virtuality the interaction between the virtual $W$-boson and the 
nucleus is most efficiently parametrized 
in terms of diffractive scattering through Pomeron exchange.  The 
Pomeron-exchange between the projectile and two or more constituent nucleons 
models the interaction between partons from different nucleons 
in the nucleus. The virtual vector boson scatters on one quark 
in the exchanged Pomeron, leading to a $Q^2$-dependence 
which is given by the $Q^2$-dependence of the Pomeron structure 
function. Thus, shadowing due to Pomeron exchange is 
a leading twist effect and survives for large $Q^2$.  
The shadowing corrections to the nuclear structure function
due to Pomeron exchange can be written as a convolution 
of the Pomeron structure function, $F_{2I\!P}$, with the Pomeron 
flux, $f_{I\!P/A}$, which describes the number density of 
the exchanged Pomerons (assuming factorization) \cite{Badelek,Melni}:  
\begin{equation}
A\delta^{(I\!P)} F_{2}^{\nu A}(x,Q^2) = \int_{ymin}^A
dy f_{I\!P/A}(y) \, F_{2 I\!P}(x_{I\!P},Q^2), 
\end{equation}
where
\begin{eqnarray}
 f_{I\!P/A} & = & - A(A-1) \gamma^2 8\pi y^{-1}
  \mbox{Re} \int_{z'>z}d^2b\, dz \,dz'\,\, \mbox{exp}[ik_L^X(z'-z)]
\\ \nonumber &\times&  \rho^{(2)}(\vec{b},z,z')\, \mbox{exp}
(-\frac{A}{2}\int_z^{z'}
\frac{d\xi}{L_X}). 
\end{eqnarray}
Here, $\gamma^2 = \sigma_{pp}/16\pi$ with $\sigma_{pp}=
21.7 s^\epsilon +56.08 s^{-\eta}$ the 
proton-proton total cross section \cite{Lands},    
$y$ is the momentum fraction of the nucleon 
carried by the Pomeron and $x_P=x/y$ is the momentum fraction of 
the Pomeron carried by the struck quark.    
$y_{min}$ is given by $y_{min}=x(1+M^2_{X_0}/Q^2)$ with  
$M^2_{X_0}=1.5$ GeV$^2$,  the minimal mass of the diffractively 
produced final states. $M_{X_0}$ is chosen such that 
it is above the relevant vector meson masses 
in order to avoid double counting. 
The mean free path of the hadronic state $X$ 
is $L_X=[\sigma_{XN}\rho(\vec{b},\xi )]^{-1}$ and we assume that the total 
cross section $\sigma_{XN}$, for the state $X$ with the nucleon, 
is independent of the mass $M_X$; we take $\sigma_{XN}=25$ $mbarn$. 
The longitudinal momentum transfer to the nucleon is 
$k_L^X=M y$.  
$ F_{2I\!P}(x_P,Q^2)$ is the structure function of the Pomeron.  
It contains a $q\bar q$ and a triple Pomeron component. 
These structure functions in Ref.\cite{Melni} have to be modified in the
neutrino induced reaction  because of the different coupling 
of the electromagnetic current 
and the weak current to the quarks in the Pomeron. 
In the $q\bar q$ component we replace 
the factor coming from the charge sum of the quarks
$(10+2\lambda_s)/9$ by the factor $4+2\lambda_s$, where the parameter 
$\lambda_s$ represents the weaker coupling of the strange quarks 
to the Pomeron compared to the $u$ and $d$ quarks; we set  
$\lambda_s \approx 0.5$.    In the triple Pomeron term we replace 
$F^{sea}_{2N}(x_{I\!P},Q^2)$ by the structure function
appropriate for W-exchange:  
\begin{eqnarray}
F_2^{(q\bar q)}(x_{I\!P},Q^2)& =&
\frac{12 (4+2\lambda_s)\beta_0^2}{\sigma_{pp}}
\frac{N_{sea} Q^2}{Q^2+Q^2_0} x_{I\!P}\,(1-x_{I\!P}),\\
F_2^{(3I\!P)}(x_{I\!P},Q^2)& =& \frac{g_{3I\!P}}{\sqrt{\sigma_{pp}}}
F_{2N}^{sea}(x_{I\!P},Q^2). 
\end{eqnarray}
The parameter $N_{sea}\approx 0.17$ (at $Q^2\sim 4$ GeV$^2$)   
is determined by the small $x$-behavior of the 
sea density \cite{Lands}. $g_{3I\!P}=0.364$ mb$^{1/2}$ and 
$\beta_0^2=3.4$ GeV$^{-2}$ are the triple Pomeron and quark-Pomeron 
coupling constants, respectively; $Q_0^2=0.485$ GeV$^2$  
is fixed by matching the photoproduction 
and deep inelastic regions \cite{Lands2}.  
The $y$-dependence of the Pomeron flux is in accordance with 
recent experimental findings by the H1 \cite{H1P} and ZEUS 
\cite{ZEUSP} Collaborations  
at Hera from diffractive deep inelastic ep scattering. These 
results also confirm that the Pomeron structure function contains 
both a hard and soft component, as had been found  by the UA8 
Collaboration \cite{UA8} previously.

\begin{figure}
\epsfig{figure=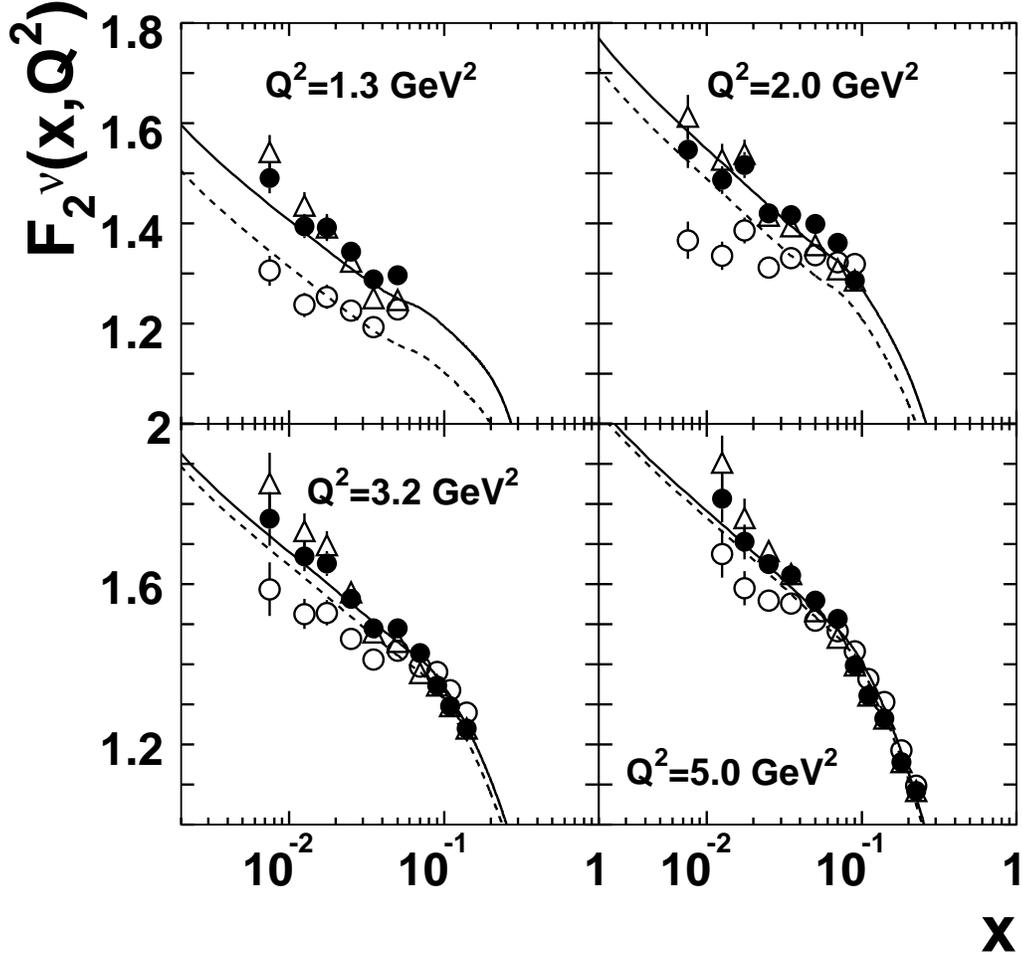,height=16.cm} 
\caption{ The CCFR data \protect\cite{CCFR} are shown as a function 
       of $x$ for different $Q^2$-values together with the parametrization 
       of Ref.\ \protect\cite{Lands2}.  
       (Data points with higher $Q^2$-values are not shown.)  
       The solid  dots show 
       the data points corrected for heavy target effects using 
       the ``two-phase'' model. The open triangles are corrected 
       using a fit to the heavy target corrections in charged lepton deep 
       inelastic scattering.  
       The open circles represent  
       the uncorrected data points. The solid and dotted curves 
       are  calculated with and without the pion contributions,  
       respectively. } 
\label{fig2} 
\end{figure}

The structure function on a heavy target $F^{\nu A}_2$  
is given in terms of the proton $F^{\nu p}_2$ 
and neutron  $F^{\nu n}_2$ structure function 
and the double scattering corrections by 
\begin{equation}
  F^{\nu A}_{2}= \frac{Z}{A} F_2^{\nu p} + 
(1-\frac{Z}{A}) F_2^{\nu n} +\delta^{(\pi)}F^{\nu A}_{2}+
\delta^{(V)}F^{\nu A}_{2}
+  \delta^{(I\!P)}F^{\nu A}_{2}. 
\label{eq:shad}
\end{equation}
In the following we will use this relation with the 
CCFR data  and the calculated 
shadowing corrections to obtain the structure function on a 
deuteron target, $F_2^{\nu D}$, which then can be compared 
to the muon structure function,  $F_2^{\mu D}$,  
on a deuteron target.  
For the discussion of the ratio $R=F^{\nu A}_2/F^{\nu D}$ 
we need a parametrization of the neutrino structure functions 
$F^{\nu p}_2$ and $F^{\nu n}_2$. 
We will use 
the parametrization of parton distributions by 
Donnachie and Landshoff 
\cite{Lands2} for small $Q^2$-values and that of the CTEQ-Collaboration 
\cite{Lai} for large $Q^2$-values. 
The parametrization of Donnachie and Landshoff 
is designed for small $Q^2$
 and matches the deep inelastic and the photo-production 
regions by taking the behavior of the structure function 
into account  for $Q^2\rightarrow 0$. This behavior is 
parametrized  by a multiplicative factor 
$(Q^2/(Q^2+Q_0^2))^{1+\epsilon}$ which models vector meson 
dominance for small $Q^2$.   
It is clear that the small $Q^2$-behavior is different
for the neutrino structure functions because of the presence of
the pion component. However, we expect that the behavior
of the non-pionic components  should be the same as in muon deep
inelastic scattering. Therefore we use the parametrization of 
Donnachie and Landshoff for the neutrino structure function 
and add the term $F_2^\pi$ in order to take into account the effects 
of PCAC.  We checked  
that this parametrization, with the pionic-term included,  
describes the CCFR data reasonably well in the small $Q^2$-region. 
In Fig.\ \ref{fig2} the CCFR data is shown as a function 
of $x$ for different $Q^2$-values together with the parametrization 
of Ref.\cite{Lands2}.  The solid dots show 
the data points corrected for heavy target effects 
according to Eq.\ (\ref{eq:shad}), while the open triangles are corrected 
using a fit to the heavy target corrections in muon deep 
inelastic scattering.  The open circles represent  
the uncorrected data points.  The solid and dotted curves 
are  calculated with and without the pion contributions, respectively.  
For small $Q^2$ the pion contributions are relatively large, but with
increasingly large $Q^2$ values the pion contributions become 
progressively less important.

As far as the
similarities and differences between
shadowing in neutrino and muon deep-inelastic scattering is concerned
we expect to see the following. In the extremely small $Q^2$-region,
where the hadronic fluctuations of the virtual photon and W-boson
dominate the structure functions (pions in the neutrino case and
vector mesons in the charged lepton case), shadowing in both
cases should be large  and should have
approximately the same magnitude. On the other hand for larger
$Q^2$-values, $Q^2 \ge 1$, 
the pion contribution becomes negligible and shadowing is
largely determined by the vector meson and the Pomeron component.
Here, we expect to see some differences
due to the different coupling of the weak current
to vector and axial vector mesons, compared to the coupling of the
electro-magnetic current to vector mesons.
  More precicely, the relative
magnitude of the VMD contribution in neutrino scattering
should be roughly half of that in the
corresponding charged lepton case.
The reason is that, although the coupling is twice as large in the
neutrino induced reaction as in the muon induced
one, the structure function is larger  by about a  factor of
$18/5$. This effect is partly compensated by  the
$A_1$, which has the same coupling to the axial current
as the $\rho^+$.  However, since the mass of $A_1$ is
large, the $A_1$ cannot account for the difference.
Note also, that the higher mass of the $A_1$ enters   in the
coherence condition for shadowing which can be important  at the
relatively low $Q^2$-values of the CCFR-data.
Finally, at large $Q^2$-values ($Q^2 > 10$ GeV$^2$) where the Pomeron
component dominates, there should be no differences in
shadowing  between neutrino and charged lepton reactions.
This is because the relative magnitude of this leading twist
component is determined by the coupling of the photon and
the W-boson to the quarks in the exchanged Pomeron.
This coupling changes in the same way as the structure functions of the
nucleons do if we go from charged lepton induced reactions to
neutrino induced ones. 
Thus,  differences in shadowing should   only occur in the higher twist
VMD terms  and should show up  in the region where shadowing of
vector mesons plays a significant role.
Since the  CCFR data have relatively small $Q^2$-values
($Q^2 \approx 1$-$15$ GeV$^2$) in the small $x$-region,
modifications of the shadowing corrections
due to vector mesons  are expected to be relevant for the CCFR-data.

\begin{figure}
\epsfig{figure=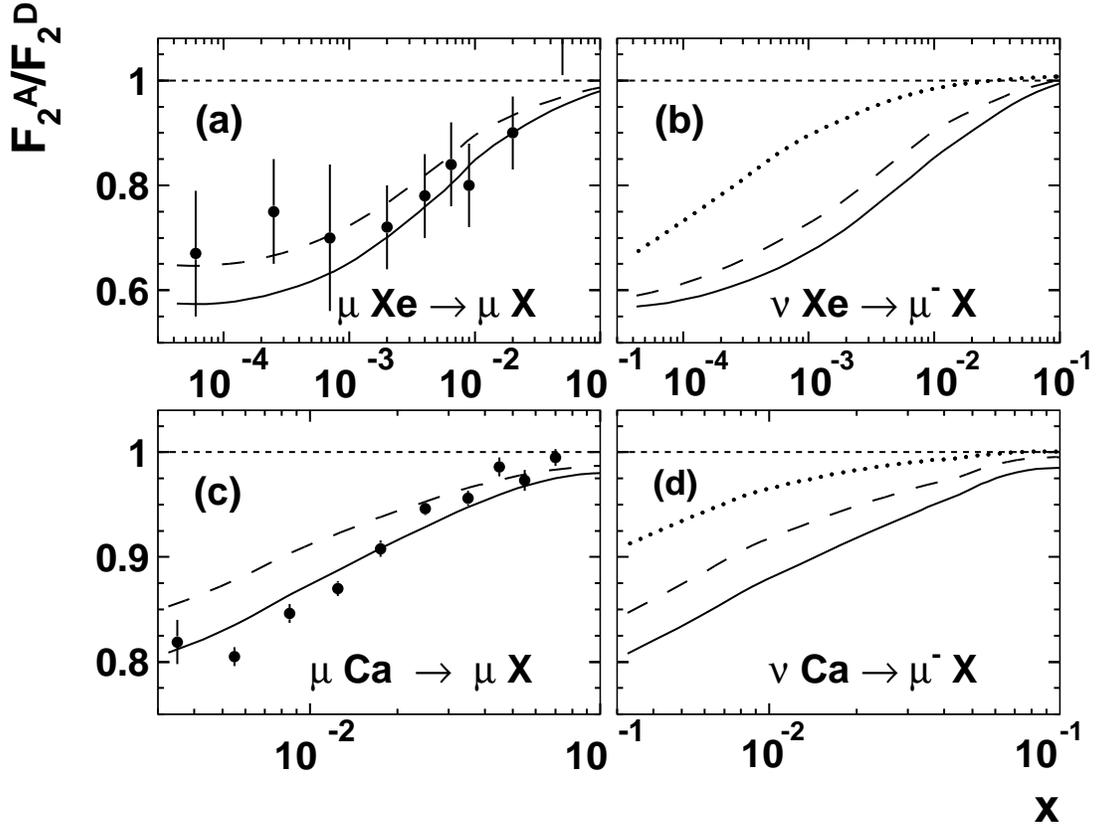,height=12.cm}
\caption{The ratios $F_{2}^{Xe}/F_{2}^{D}$  and
           $F_{2}^{Fe}/F_{2}^{D}$ are   calculated in the two phase
         model for charged lepton (a,c)
         and neutrino  deep inelastic scattering (b,d).
         The dotted and the dashed  curves stand for
         shadowing due to the PCAC component  alone,
         and  due to PCAC with VMD contributions
         also included. The solid curve is the total shadowing.
         The data are for muon scattering from Ref.[35,36].
         For the structure functions we used the parametrization of
         Ref.[31]. }
\label{fig3}
\end{figure}

In order to highlight the similarities and differences
between shadowing in charged lepton and neutrino scattering, we calculated
the shadowing  corrections to the structure functions for
both reactions.
Since there are experimental data for
the ratio between the structure functions of Xenon/deuteron
and Ca/deuteron measured for charged lepton scattering by the
E665-collaboration \cite{E665} and by the NMC-Collaboration \cite{NMCs},
we calculated these ratios for both charged lepton and neutrino scattering.
The results for the muon induced reaction and their comparison
with the experimental data can be found in Ref.\cite{Melni}.
Here, we show them in Fig.\ref{fig3}a and b
for comparison.
While the dashed curves stand for shadowing calculated only with
vector mesons, the solid curves also include the Pomeron contributions.
We stress that the
points are calculated for the $x$ and $Q^2$-values of the experimental
data-points. They are shifted in $x$ for clarity. The experimental data
are represented by solid dots with statistical and systematic errors
added in quadrature.
We note that the calculation
describes the experimental data reasonable well and that
the important contribution to shadowing comes from VMD in the muon case.

The shadowing corrections in  neutrino deep inelastic
scattering are  shown in Fig.\ref{fig3}b for $Xe$ and in Fig.\ref{fig3}d
for $Ca$, respectively.
Here, the dotted curves are the results with only pion contributions,
the dashed with pion and vector meson contributions, and the
solid curves include the Pomeron component also and describe
thus the total shadowing.
We see that the total  shadowing in the neutrino induced reaction is
comparable in magnitude
to shadowing in the charged lepton induced reactions.
However, the relative importance of the individual contributions to shadowing
are very much different.
While  the PCAC  term dominates
in the small $x$-region, the VMD and Pomeron contributions
become more and more important with increasing $x$ (which,
in these experiments, is correlated with increasing $Q^2$)
and shadowing is largely determined by
their interplay.

\begin{figure}
\epsfig{figure=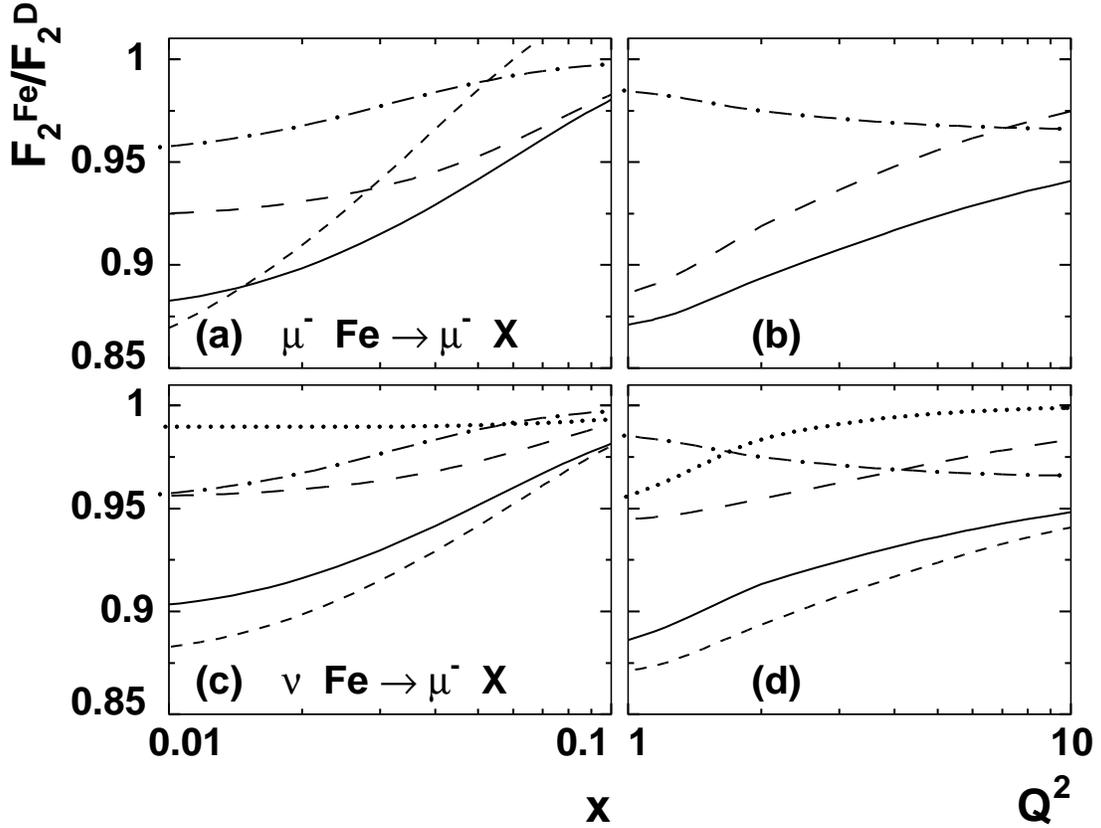,height=12.cm}
\caption{The different contributions to shadowing on $Fe$ in muon ((a) and
          (b)) and
         neutrino ((c) and (d))
         deep inelastic scattering as a function of
         $x$ for fixed $Q^2=3$ GeV$^2$ ((a) and (c)) and as a function of
         $Q^2$ for fixed $x=0.02$ ((b) and (d)).  The dotted, dashed  and
         dash-dotted curves stand for the pion, VMD and Pomeron
         contributions, respectively. The total shadowing 
         corrections  in the 
         muon induced reaction are shown  as the 
         short dashed curves in (c) and (d)  for comparison.
         The ``$Q^2$ independent'' fit is also shown  
         as short dashed curve in (a) for
         comparison.}
\label{fig4}
\end{figure}

Next, we focus on the effects arising from the differences in VMD
between the neutrino and the charged lepton case and their relevance
to the CCFR data. We calculated the relative contributions of the different
components to shadowing on an iron target
in the kinematical region of the CCFR experiment.
The result is shown in Fig.\ref{fig4},
where the ratio, $R=F_2^{Fe}/F_2^{D}$,
is plotted as a function of $x$ for fixed $Q^2=3$ GeV$^2$ and
as function of $Q^2$ for fixed $x=0.02$ for neutrino and
charged lepton scattering.
We see that leading twist shadowing (Pomeron component) is the same for both
neutrino and charged lepton induced
reactions and is important for high $Q^2$-values (dash dotted lines).
Further, shadowing due to vector mesons
is much more significant for charged lepton
than for neutrino  deep inelastic scattering (dashed lines).
In the former it plays
an important role even at relatively high $Q^2$. Note also
that the pion component is negligible  above $Q^2>3$ GeV$^2$.
For comparison the ``$Q^2$ independent'' shadowing
is shown in Fig.\ref{fig4}a  (short dashed line).
We see that shadowing for {\it fixed}
$Q^2$ is not to be described by such a parametrization.
This shows the strong $Q^2$-dependence of shadowing  in the
available charged lepton data. 

\begin{figure}
\epsfig{figure=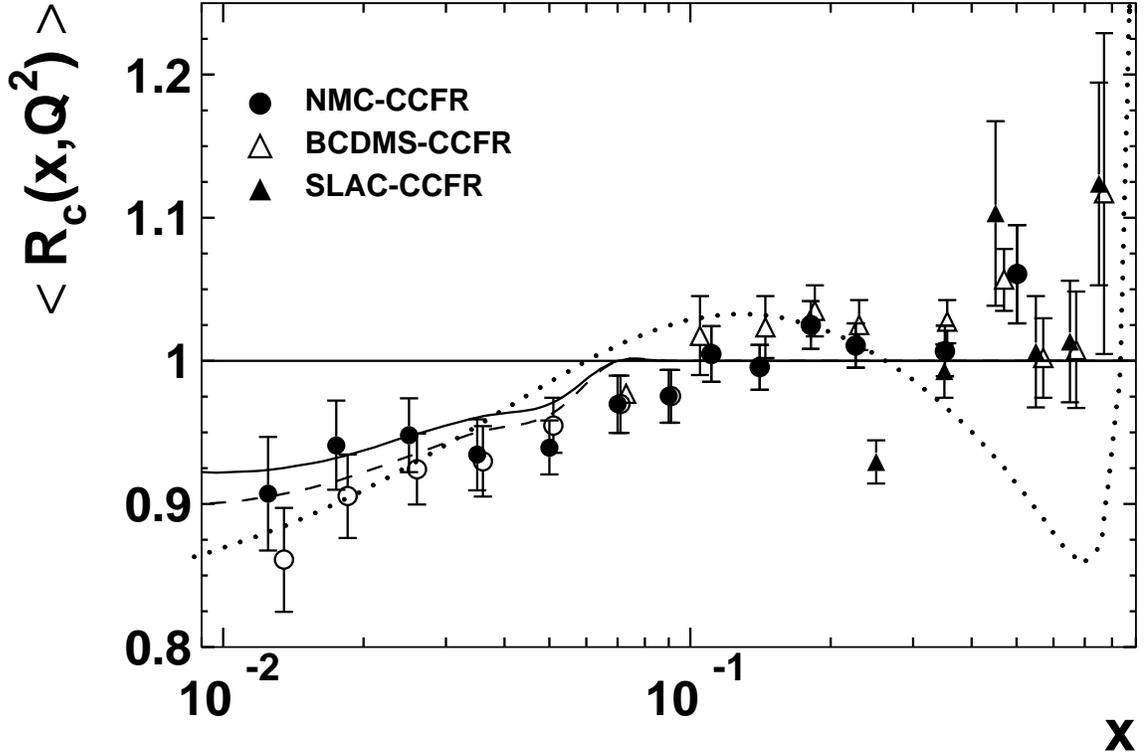,height=12.cm}
\caption{The charge ratio as a function of $x$  calculated using
         the CCFR [1] data for neutrino and
         NMC [2], SLAC [37] and BCDMS [38] data for muon
         induced structure functions. The data have been integrated
         above $Q^2=2.5$ GeV$^2$ over the overlapping kinematical regions and
         the statistic and systematical errors are added in
         quadrature.
         The heavy target corrections are calculated by using the
         ``two phase-model''
         in the shadowing region
         and a fit to the experimental data on nuclear shadowing
         in the non-shadowing region (black circles) and by using
         the $Q^2$ independent fit in the entire region (open circles).
         The ratio $R=F_2^{Fe}/F_2^D$ calculated for neutrino
         and for charged lepton scattering,
          is shown as solid  and dashed lines,
         respectively. They are calculated in the ``two phase model''
         and are averaged over the same $Q^2$-regions as the data.
         The $Q^2$ independent fit is represented by a
         dotted line.  }
\label{fig5}
\end{figure}

Having seen how shadowing in muon deep inelastic scattering
compares with shadowing in neutrino deep inelastic scattering
we calculate the shadowing corrections for
the CCFR-data on an iron target. We apply these corrections for each data
point and integrate over $Q^2$ (above $Q^2=$2.5 GeV$^2$)
 where the CCFR-data \cite{CCFR}
and the NMC-data \cite{NMC}
overlap (in order to obtain better statistics) and calculate the
``charge ratio''.
In the non-shadowing region ($x\ge 0.07$) we use the
$Q^2$-independent  parametrization
of $F_{2}^A/F_{2}^D$ measured in charged lepton induced processes.
 The result is shown in Fig.\ref{fig5}
(black circles). The statistical and systematic errors are added
in quadrature.
The result we would have if
we used the $Q^2$-independent parametrization of the muon
shadowing data  in  the
shadowing region is shown as open circles for comparison.
The shadowing correction factors
are shown as solid and dotted lines for the ``two-phase'' model and
the ``$Q^2$-independent'' shadowing, respectively.
The shadowing correction for charged lepton  scattering calculated
in the ``two-phase'' model is shown as dashed line for comparison.
These ratios $R=F_2^{Fe}/F_2^{D}$
have been obtained according to Eq.(3.17).
While in the neutrino case we used the data for $F_2^{Fe}$ together
with the corrections ($\delta F_2^\pi$ ...) to calculate $F_2^D$ and
the ratio $R$,
in the charged lepton case we used a parametrization for $F_2^D$
and the shadowing corrections to calculate $F_2^{Fe}$ and thus the ratio
$R$.
Since the data include points with relatively high $Q^2$-values
we use the parametrization of Ref.\cite{Lai}
(CTEQ4L) for the parton distributions.
We also include the data from SLAC \cite{SLAC} and BCDMS \cite{BCDMS}
for completeness. 

The differences between the  calculated and
``fitted'' shadowing corrections are partly due to
the difference between shadowing in neutrino and muon scattering
and partly due to the $Q^2$-dependence of shadowing, as can be seen
in Fig.\ref{fig5}.
In connection with the $Q^2$-dependence we note that
the parametrization of the shadowing corrections has been
obtained by fitting the ratio $R=F_2^{\ell A}/F_2^{\ell D}$, 
in charged lepton deep inelastic scattering. 
In the small $x$-region
this fit is mainly determined by the NMC data on Ca \cite{NMCs}.
However, the NMC-data for the structure function ratio
 have lower $Q^2$-values in the first  $x$-bins
than the CCFR-data we use to calculate the charge ratio.
The average $Q^2$ for the NMC ratio $R=F_2^{Ca}/F_2^{D}$
are $Q^2=1.9$, $2.5$, $3.4$, $4.7$ GeV$^2$ for  $x=0.0125$, $0.0175$ ,
$0.025$, $0.035$, respectively.
On the other hand  we integrate the CCFR-data
above $Q^2=3.2$ GeV$^2$ and have $Q^2=4.1$, $5.5$, $7.9$, $9.7$ GeV$^2$
for the averaged $Q^2$ values for the same $x$ bins.
Since VMD  is more important for lower $Q^2$, it is clear that
the parametrization of the NMC-data overestimates the shadowing.

\section{Extraction of the strange quark distribution} 

Now that we have determined the shadowing corrections for 
neutrinos, we can examine  
how they influence the determination of strange quark densities.  
Currently there are two viable  
methods for the extraction of strange quark parton distributions.  
The ``direct'' method utilizes charm-hadron production in neutrino 
deep-inelastic scattering. The triggering signal for this process 
is the measurement of opposite sign dimuons, one coming from the 
lepton vertex, while the other comes from the semi-leptonic decay of 
the charmed hadron \cite{CCFRLO,CCFRstr}. The other method is to obtain 
the strange quark distribution by comparing charged lepton deep 
inelastic scattering with neutrino deep inelastic scattering. In the 
second case the strange quark distribution can be extracted from 
the relation 
\begin{equation}
 \frac{5}{6}F_2^{\nu N_0}(x,Q^2)-3F_2^{\mu N_0}(x,Q^2) =xs(x)+\frac{x}{3} 
[s(x)-\bar s(x)]. 
\label{eq:ssbar} 
\end{equation}
Eq.\ \ref{eq:ssbar} follows if we assume parton charge symmetry and
neglect charm quark contributions.  
If one assumes that $s(x)=\bar s(x)$, the difference between the
neutrino and muon structure functions 
measures the strange quark distribution in the nucleon. 
Experimentally, the two methods for determining the strange quark
distribution are not compatible in the region of small $x$. 
This conflict is also reflected in the fact that the  
``charge ratio'' $R_c$ is different from one in this region.  
If we had used the ``correct'' strange quark distribution the 
charge ratio would be unity (assuming $s(x)=\bar s(x)$).   

\begin{figure}
\epsfig{figure=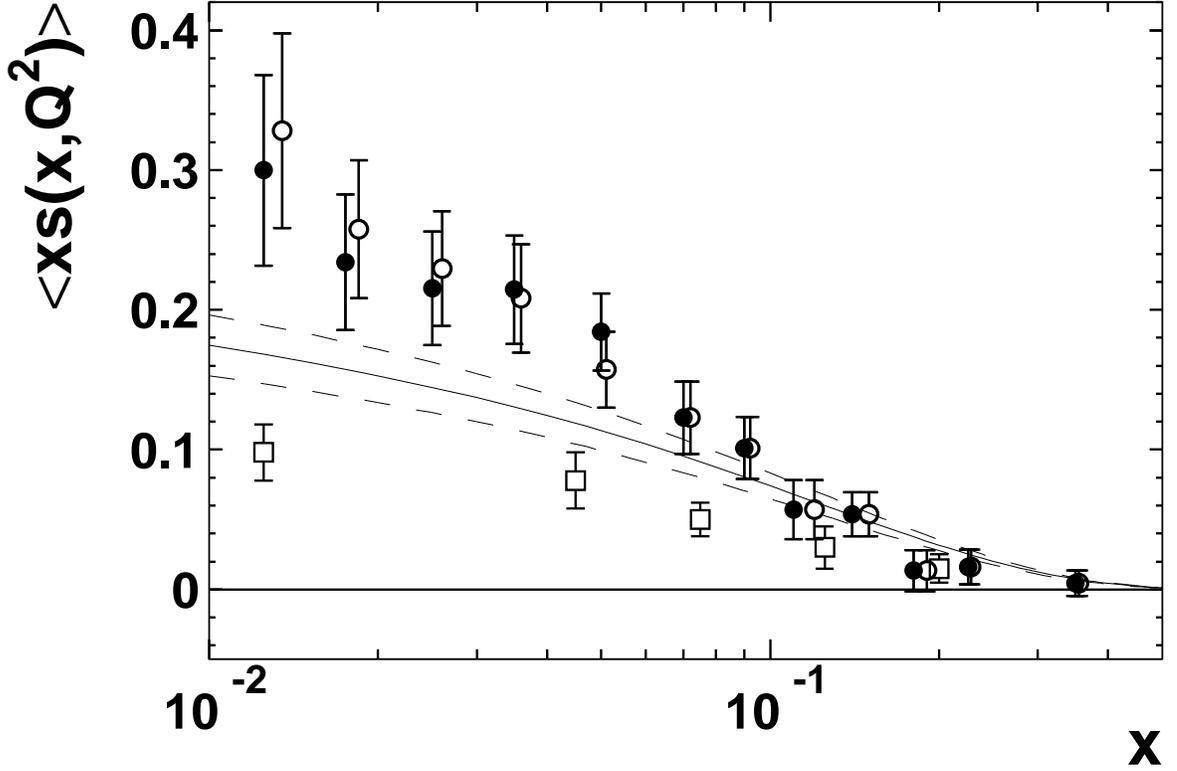,height=14.cm}
\caption{The  strange quark distribution extracted
         from the CCFR and NMC data assuming the validity of
        charge symmetry and $s(x)=\bar s(x)$.
        The data have been integrated over the overlapping
        $Q^2$ region to obtain better statistics.
        The solid (open)   circles stand for
        $5/6 F_2^{\nu}- 3 F_2^\mu$ using the two phase model
       (using the $Q^2$-independent parametrization)
        for the shadowing corrections.
        The open boxes stand for the LO CCFR determination of the strange
        quark density from dimuon production at $Q^2=4$ GeV$^2$
         [39].
        The solid line is the NLO CCFR determination at $Q^2=4$ GeV$^2$
          [40].
         The band around the NLO curve indicates the
        $\pm 1\sigma$ uncertainty in the distribution.
        }
\label{fig6}
\end{figure}

We converted the CCFR neutrino data on iron to deuteron data by 
applying our shadowing corrections.  We then extracted 
the strange quark distribution according to Eq.\ \ref{eq:ssbar}. 
In order to get better statistics we integrated the structure functions 
over the overlapping $Q^2$-regions, as before.  
The result  is shown in Fig.\ \ref{fig6}, where the strange 
quark distributions extracted with 
the ``two-phase'' shadowing and the ``$Q^2$-independent''  
shadowing corrections are  shown as black and open circles, respectively.    
Statistical and systematic errors are added in quadrature. 
The strange quark distribution 
as determined by the CCFR Collaboration  
in dimuon production using a LO analysis \cite{CCFRLO} is shown 
as open boxes, while the distribution extracted in 
NLO analysis \cite{CCFRstr} 
from dimuon data is shown as a solid line. The band around the NLO curve 
indicates the $\pm 1\sigma$ uncertainty.  
Although the strange quark distribution obtained from the difference 
between the neutrino and muon structure functions using the ``two phase''  
model for shadowing is smaller in the small $x$-region than that obtained 
by applying the $Q^2$-independent shadowing, both distributions 
are incompatible with the strange quark distribution extracted from 
dimuon production.  

\begin{figure}
\epsfig{figure=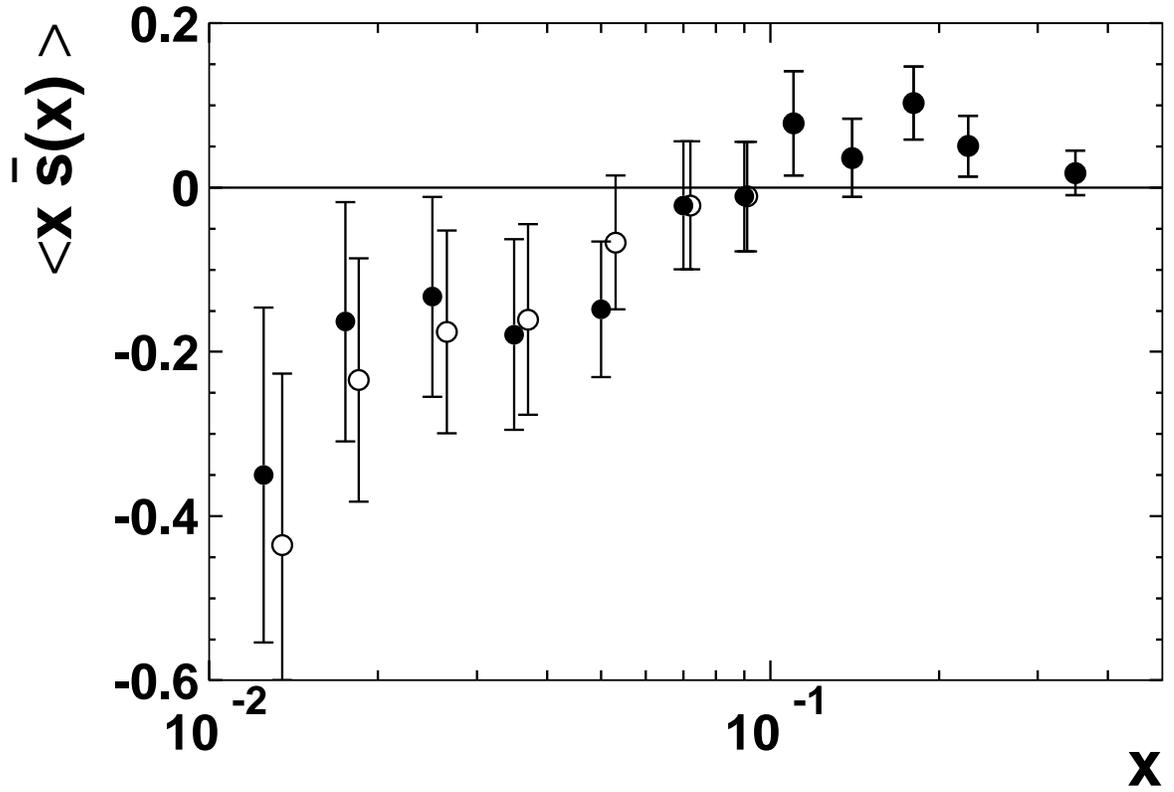,height=14.cm}
\caption{The (physically unacceptable) 
 anti-strange quark distribution  extracted from the data
   assuming that the discrepancy between the muon and neutrino structure
   function is due to different strange quark and anti-strange quark
   distributions and that the strange quark distribution is given
   by that extracted from  di-muon experiments. }
\label{fig7}
\end{figure}

The remaining discrepancy could be attributed to {\it different} 
strange and anti-strange quark distributions \cite{Brodsky,sdiff} 
in the nucleon. From Eqs.\ (\ref{Rc}) and (\ref{eq:ssbar}) and 
Fig.\ 5, we see that 
the difference $s(x)-\bar s(x)$ should be positive for small 
$x$-values ($x<0.1$). This is in contradiction with  
the analysis of Ref.  
\cite{Brodsky} but agrees qualitatively with that in Ref.\cite{sdiff}.  
Note in this connection that the experimentally determined 
structure function, $F_2^{CCFR}$, is a flux weighted average of the neutrino
and anti neutrino structure functions \cite{CCFR}. 
Since neutrino events dominate 
over the anti neutrino events in  the event sample of the CCFR
experiment, it can be approximately regarded as neutrino
structure function. 
In Fig.\ \ref{fig7} we extract the strange antiquark distribution 
vs.\ $x$ using Eq.\ (\ref{eq:ssbar}).  We use the experimental data 
for the muon and neutrino structure functions (with our calculated
shadowing corrections), together with the strange quark distribution
measured in dimuon production.  Note that with this method   
we obtain a {\it negative} strange antiquark distribution for
small $x$-values!  This strongly suggests that the entire discrepancy 
cannot be attributed to the difference between $s(x)$ and $\bar s(x)$. 

\section{Conclusions} 

In conclusion, we have carefully re-examined shadowing corrections to 
the structure function $F_2^\nu$ in deep inelastic neutrino 
scattering on an iron target. Although the shadowing corrections are not 
as large as one would naively expect, they are still 
sizable and similar to shadowing in charged lepton induced 
reactions in the small-$x$ region. Taking neutrino 
shadowing corrections into account properly  
resolves part of the discrepancy between the 
CCFR neutrino and the NMC muon data in the small $x$-region.  
Neutrino shadowing corrections also remove part of 
the corresponding discrepancy between the 
two different determinations of the strange 
quark densities.  However, the charge ratio $R_c$, of Eq.\ 
\ref{Rc}, still deviates from unity at small $x$.  
Furthermore, the 
data rules out the possibility that the discrepancy is entirely 
due to the difference between the strange and anti-strange 
quark distributions.  
We are therefore forced to consider the possibility of a rather 
uncomfortably large charge symmetry violation in the sea quark 
distributions.   This will be discussed in a subsequent paper 
\cite{Bor98}.

\acknowledgments 

We would like to thank S. Brodsky, 
 B.Z. Kopeliovich and W. Melnitchouk for helpful  
discussions. This work is supported by the Australian  
Research Council and by the National Science Foundation under contract 
NSF-PHY-9722706.

\end{document}